\documentclass[20pt,a4paper,twoside]{article}
\usepackage{epsfig}
\usepackage{baltlat6}
\usepackage{wrapfig}
\begin{document}
\ \
\vspace{-0.5mm}

\setcounter{page}{91}
\vspace{-2mm}

\titlehead{Baltic Astronomy, vol.\,20, 91--106, 2011}

\titleb{DETERMINATION OF HOMOGENIZED EFFECTIVE\\
 TEMPERATURES FROM STELLAR CATALOGS }

\begin{authorl}
\authorb{V.~Malyuto}{1}
and
\authorb{T.~Shvelidze}{2}
\end{authorl}

\begin{addressl}
\addressb{1}{Tartu Observatory, T\~oravere, 61062, Estonia;
valeri@aai.ee}
\addressb{2}{Abastumani Astrophysical Observatory, Kazbegi ave. 2a,
0160, Tbilisi, Georgia}
\end{addressl}

\submitb{Received:  2010 July 8; revised: 2011 March 8; accepted: 2011
March 21}

\begin{summary} Some selected catalogs of the effective temperatures
($T_{\rm eff}$) for F, G and K stars are analyzed.  By an improved
technique we estimate the external errors of these catalogs from data
intercomparisons.  The $T_{\rm eff}$ values are then averaged with the
appropriate weights to produce a mean homogeneous catalog based on the
selected data.  This catalog, containing 800 stars, is compared with
some other independent catalogs for estimating their external errors.
The data may be used as a source of reliable homogeneous values of
$T_{\rm eff}$, together with their errors.  \end{summary}


\resthead{Homogenized effective temperatures from stellar catalogs}
{V. Malyuto, T. Shvelidze}

\sectionb{1}{INTRODUCTION}

Many published stellar catalogs are available which contain the main
astrophysical parameters of stars ($T_{\rm eff}$, $\log g$, [Fe/H]), and
these data are in use for investigations of the Galaxy structure, its
star formation and history of chemical enrichment.  Such catalogs are
based on the various observational data obtained by different methods.
Evidently, the errors of astrophysical parameters in various catalogs
(sometimes in the same catalog) are different and frequently are known
with insufficient accuracy.  However, for application of the catalog
data, as well as for creation of the weighted compiled catalogs, we need
reliable errors.  As it was noted in a recent paper of Soubiran et al.
(2010), serious efforts should be undertaken to create extensive and
homogeneous catalogs of $T_{\rm eff}$, $\log g$ and [Fe/H] covering the
whole HR diagram and metallicity range.  Such homogeneous catalogs of
reliable astrophysical parameters can be also used to select lists of
stars applicable for the data calibration in spectral or photometric
surveys in the Galaxy.

Compiling the catalogs of stellar parameters, the authors usually begin
with reducing all the data sources to a system of one reference catalog,
however, next steps in many cases are different.  In two recent papers
Taylor (2005) and Borkova \& Marsakov (2005) have produced large
compiled catalogs of the published [Fe/H] values.  In both cases similar
approaches have been used:  the resulting mean [Fe/H] values were
obtained by averaging the data with weights derived from the residuals.
Such a procedure may be effective only in the case when three or more
values of the parameter for the same star are available.

For the catalogs with stars in common, Cenarro et al.  (2007) have
calculated the variances of data differences with the reference catalog
what allows to assign some weights to each catalog and to compile the
weighted data for $T_{\rm eff}$, $\log g$ and [Fe/H].  However some
involved catalogs may be heterogeneous in accuracy, and this approach
does not allow to estimate the external errors for individual catalogs
which are important for many applications.  Fortunately, when the error
of one of the selected catalogs is known, the external error of another
homogeneous independent catalog may be estimated from the data
intercomparisons, and this approach has been used by some authors (see,
e.g., Ducourant et al. 2006 for proper motions).

More productive approaches may be used when triples of homogeneous
catalogs with the stars in common are taken together and the external
errors of some catalogs are estimated from the data intercomparisons
(some early examples of such approach may be found in Chun \& Freeman
(1978) for $B$,$V$ photometry and in Evans et al.  (2002) for proper
motions).  In this approach, Malyuto (1994) has applied a special
technique for metallicities, where in every triple of catalogs some
appropriate weights are assigned to variances of data differences in
each pair of the catalogs (for the stars in common to all three
catalogs, and for the stars in common only to two catalogs, separately).
In our Paper I (Malyuto \& Shvelidze 2008) this technique was used to
estimate the external errors of $T_{\rm eff}$ for five independent
catalogs where all possible triples were considered.

In the present paper we re-analyze the data from the same five catalogs
but with the use of an improved technique, which determines external
errors of the catalogs by solving linear equations (with variances of
the data differences as the function of errors) by the least squares,
when the number of catalogs with the stars in common exceeds three.  We
also define some new improved homogeneous subsamples in these catalogs
and estimate their external errors from pairs assuming that the error of
one considered catalog in every pair is already obtained as described
above.  Then trying to produce the mean homogenized catalog we average
the values of $T_{\rm eff}$ for every star from two or more catalogs
with the weights which are inversely proportional to the squared errors.
Possible extension of this catalog with the use of some additional data
is discussed.

\sectionb{2}{GENERAL PRINCIPLES USED IN THE COMPILING OF CATALOGS}

Let $X_i$ denote a quantity in the $i$-th catalog (measured without
bias).  If we like to obtain a good estimate of the quantity being
measured, we should consider weights, $w_i$, in the following sum:
$$w_1X_1+ ...+w_nX_n, i=1,2, ...,n, $$
where  $n$ is the number of used catalogs.
A simple statistical  analysis as described in textbooks shows that the
weights should be chosen  to be inversely proportional to the  errors
squared:
   \begin{equation}
w_1={1\over {{\sigma_1}^2}} \hspace{1cm} \mbox{...} \hspace{1cm}  w_n={1\over {{\sigma_n}^2}}.
   \end{equation}
Then a good estimate of the quantity (weighted mean) is
   \begin{equation}
\overline{X}={{\sum X_i w_i}\over {\sum w_i}}.
   \end{equation}

A standard error of the weighted mean is equal to
\begin{equation}
\sigma_{\overline{X}}=\sqrt{1\over {\sum w_i}}.
   \end{equation}

In the case of catalogs of stellar astrophysical parameters, many
published errors $\sigma$ are unknown or uncertain.  Therefore a direct
simple approach to averaging the catalogs using their published errors
does not prove to be efficient in many cases.

However, there is a possibility to estimate primarily the errors of the
catalogs through data intercomparisons (if the used catalogs are
homogeneous and statistically independent).  Thus we should consider
variances of data differences for every pair of the catalogs.  Say, for
Catalogs 1 and 2 they are:
\begin{equation}{{\delta^{2}_{12}}}={{{{{\sum_{j=1}^{N}}}{{({X_{1,j}-X_{2,j}})}^2}}\over {N}-1}},
   \end{equation}
where  $N$ is the number of stars in common; other
analogous quantities  (${\delta^{2}_{13}}$ and so on) should be
calculated  for all the catalogs. The $3\sigma$ rule is applied to the
data to reject the stars which may be considered as outliers.
From the rule of addition of variances we may write:
   \begin{equation}
 {\delta^{2}_{12}}={\sigma_1}^2+{\sigma_2}^2   \end{equation}
   \begin{equation} {\delta^{2}_{13}}={\sigma_1}^2+{\sigma_3}^2
   \end{equation}   \begin{equation}
 {\delta^{2}_{23}}={\sigma_2}^2+{\sigma_3}^2
   \end{equation}
for Catalogs 1, 2, 3, etc.  If there are three catalogs with some stars
in common, the errors may be derived from these variances of the data
differences.  If there are more than three catalogs with some stars in
common, we solve these linear equations by the least squares to derive
the errors of the catalogs.

Another possibility to derive the errors is to use only the pairs of
catalogs, assuming that the error $\sigma_1$ of Catalog 1
(considered as the basic catalog) is known in advance (it might have
already been determined using the approach described above for some
selected catalogs).  Then the errors of Catalogs 2, 3 and so on may be
calculated from the appropriate variances of the data differences in the
form:
\begin{equation}
\sigma_2^2={{\delta_{12}}^2-{\sigma_1}^2}
\end{equation}
\begin{equation}
\sigma_3^2={{\delta_{13}}^2-{\sigma_1}^2}
\end{equation}

Finally, the obtained errors of the catalogs may be inserted into
Eq. (1) to obtain the weights and to produce a mean compiled
homogenized catalog of the parameters with the use of these weights in
Eqs. (2) and (3).

\sectionb{3}{ANALYSIS OF THE SIMULATED CATALOGS}

A question may arise, whether the treatment of some catalogs
simultaneously (up to five in our case) improves the errors of the
catalogs obtained from the data intercomparisons.  To verify this, we
took some simulated catalogs with the selected ${\sigma T_{\rm eff}}$
values and considered
the results, obtained with the use of our technique, for
estimating errors described in the previous Section.  The simulated data
were produced with random variables drawn from the Gaussian distribution
for a given value of $T_{\rm eff}$ = 6000 K and for a given ${\sigma
T_{\rm eff}}$ (30, 50, 60 or 70 K in different combinations), with
numbers of stars in common $N$ = 50, 100, 200, 500, respectively.  The
results are presented in Table 1. Two possibilities were considered:
when the given errors in every combination are the same (see the first
part of Table) and they are different (the second part of Table).


\vspace{3mm}
\begin{table}
\parbox{120mm}{{\smallbf\ \  Table 1.}\small\baselineskip=9pt\  Differences of
${\sigma T_{\rm eff}}$ (obtained minus given ${\sigma T_{\rm eff}}$
values) for some combinations of the selected simulated  catalogs
(the combinations are given in the first column).  Triples,
quadruples and quintuples of the simulated catalogs are considered.}
\begin{center}
\vbox{\footnotesize
\begin{tabular}{llllll}
\tablerule
Given ${\sigma T_{\rm eff}}$ \hfill &$N$=50 \hfill \hfill &$N$=100 \hfill  &$N$=200 \hfill &  $N$=500 \\
\tablerule
30/30/30          &-8/7/7 &   -4/2/2  & -4/4/5   &  1/0/1      \\
30/30/30/30   &8/-2/-5/-7    & 0/-3/5/0  & -3/3/2/0 & 0/2/1/0 \\
30/30/30/30/30   &6/-3/-5/-3/5 & -1/-4/6/0/1  & -2/1/3/0/-2 & -1/3/0/-2 \\
50/50/50          &-13/11/12&  -7/3/4  & -6/6/9   &  0/1/2     \\
50/50/50/50   &14/-4/-8/-11  & 0/-5/8/-1  & -5/5/3/-1 & 0/3/-1/0 \\
50/50/50/50/50   &11/-4/-8/-6/9 & -1/-6/10/-1/2 & -3/1/4/0/-3 &-3/1/4/0/-3  \\
70/70/70         &-18/16/16 &-10/5/5& -8/8/12  & 3/1/1       \\
70/70/70/70   &19/-6/-11/-15 & 0/0/2/12 & -1/5/10/-1 & 2/2/1/2 \\
70/70/70/70/70  & 15/-6/-12/-8/12& -2/-9/14/-1/3 & -4/2/6/0/-4 & -1/2/0/0/-3 \\
\tablerule
50/70/70          &-17/17/15 &-12/5/5& -13/8/12 &  2/1/1     \\
50/50/70         &-14/11/14  &-9/5/4 & -9/8/10  &  2/0/1      \\
30/70/70          &-22/18/14 & -19/5/5& --/8/12   &  2/1/1    \\
30/50/70         &-14/13/13  & -13/5/4& -19/8/11 &  2/0/1     \\
30/30/70         &-10/8/12   & -9/6/3  & -10/8/9   & 2/0/2       \\
30/50/50/70   & -3/4/15/-8   & -1/0/3/6 & -1/4/7/0 & 1/2/0/2 \\
50/60/60/70   & 8/6/17/-10   &  0/0/2/9 & -1/4/8/-1 & 1/2/0/2 \\
30/50/50/60/70   & 12/-7/-11/-4/13& -7/-7/10/0/-2 & -8/0/6/1/-2 & 0/1/0/0/-2 \\
50/50/60/70/70   & 14/-9/-2/-8/15 & -4/-7/12/-1/3 & -6/2/6/0/-4 & -1/2/0/0/-2 \\
\tablerule
\end{tabular}
}
\end{center}
\vspace{-5mm}
\end{table}

This Table shows that, in general, the differences become smaller (the
results more reliable) with increasing number of stars in common (from
50 to 500), as well as with increasing number of the involved catalogs
(from 3 to 5).  The differences are the largest when one considerably
deviating value is present (say, 30 K/70 K/70 K).  For just mentioned
combination of catalogs we receive an uncertain result.  In the present
analysis with real data (see below), the numbers of stars in common are
up to several hundreds and exceed 30 for every pair in the compared
catalogs.

\sectionb{4}{SELECTED CATALOGS}

The selected catalogs of F, G and K stars with $T_{\rm eff}$ values are
the same as in Paper I, they are presented in the list given below.  The
catalogs were selected to be independent and as abundant as possible.

1. Catalog 1 of Masana et al.  (2006) with 10999 F, G and K dwarfs and
giants, values of $T_{\rm eff}$ were obtained using the Spectral Energy
Distribution Fit method, the data were taken from 2MASS ({\it
JHK}$_s$) photometry, the [Fe/H]) and $\log g$ data are also available
(Masana 2008).

2. Catalog 2 (Valenti \& Fischer 2005) is a catalog of stellar
properties for 1040 nearby F, G and K dwarfs and giants based on fitting
of the observed spectral energy distributions and synthetic spectra.

3. Catalog 3 (Edvardsson et al. 1993) contains 189 nearby F and
G dwarfs with effective temperatures derived from $b-y$
photometry calibrated with a grid of synthetic colors.

4. Catalog 4 (Kovtyukh 2011) contains the determinations of $T_{\rm
eff}$ for 647 F, G and K dwarfs and giants from the line depth ratios
measured in high resolution spectra (collected with the ELODIE
spectrometer at the Haute-Provence Observatory) using calibrations from
different sources.

5. Catalog 5 of Ramirez \& Melendez (2005) for 754 dwarfs and giants in
which for determining $T_{\rm eff}$ the Infrared Flux Method (IRFM) has
been applied with the use of IR photometry.

Catalog 1, the largest one, is considered as the reference catalog.  In
Figure 1 of Paper I, we considered the systematic trends of $T_{\rm
eff}$ differences between Catalog 1 and other catalogs of this list as a
function of three main astrophysical parameters ($T_{\rm eff}$, $\log
g$, [Fe/H]).  It was shown that the $T_{\rm eff}$ differences are more
scattered for cooler stars and for stars with [Fe/H]\,$<$\,--1.1.  To
deal with homogeneous data, we decided to use in the present analysis
only the stars with the $T_{\rm eff}$ values within 5200--6700~K and
[Fe/H]\,$>$\,--1.1.

In the present paper, the values of $T_{\rm eff}$ from Catalogs 2--5
were compared with the values from Catalog 1 using the stars
irrespective of their $T_{\rm eff}$ but only with [Fe/H]\,$>$\,--1.1 to
check if these catalogs show systematic effects.  No effects were found
for Catalogs 3 and 4, but some small trends are present for Catalogs 2
and 5 (Figure 1).  The temperatures from these catalogs were transformed
by linear equations to the system of Catalog 1. Only the transformed
data are used in the following treatment.

\begin{figure}[!t]
\vbox{
\centerline{\psfig{figure=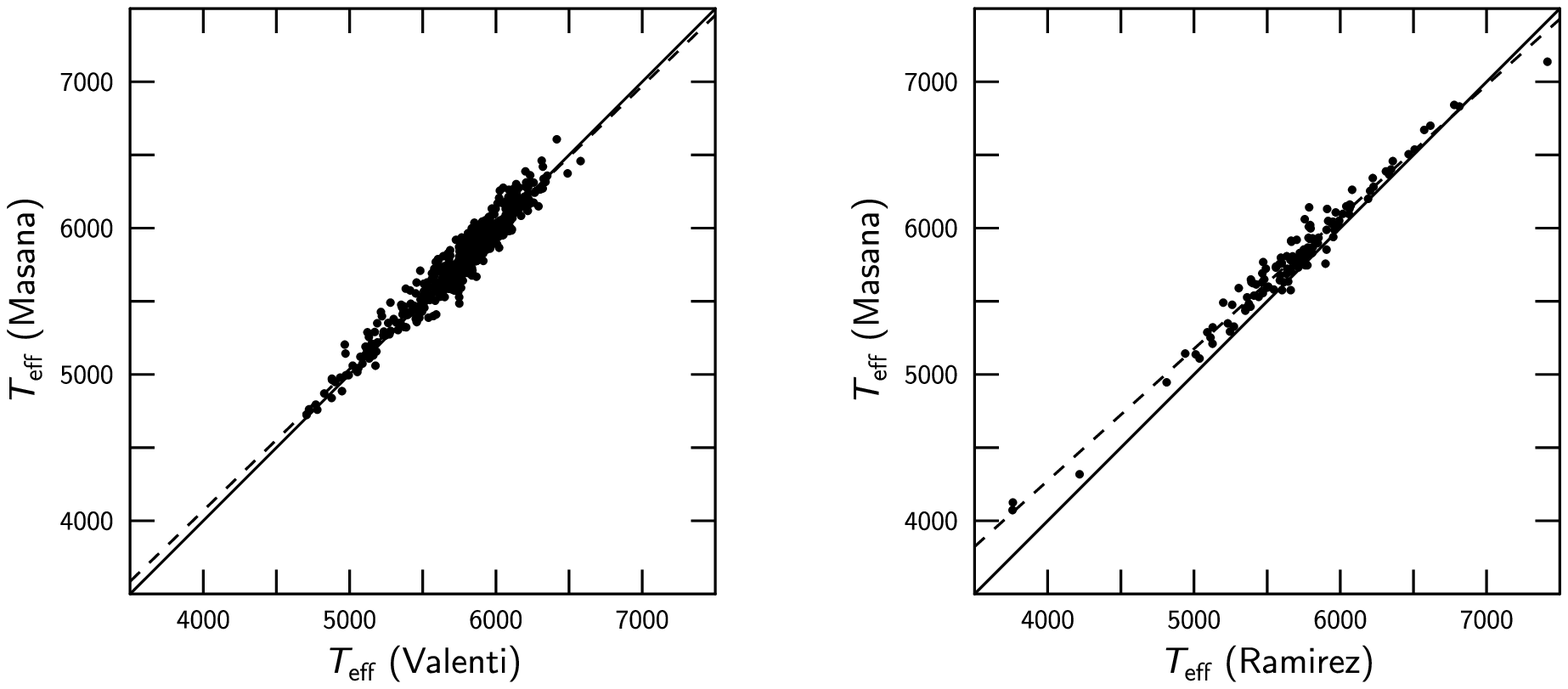,width=124truemm,angle=0,clip=}}
\captionb{1}
{Comparison of Catalogs 2 (Valenti \& Fisher 2005) and 5 (Ramirez \&
Melendez 2005) with Catalog 1 (Masana et al. 2006). The linear
transformation equations (broken lines) were applied
for reducing Catalogs
2 and 5 to the system of Catalog 1:
$T_{\rm eff}1$ = 0.9672\,$\times$\,$T_{\rm eff}2$ + 201 and
$T_{\rm eff}1$ = 0.9014\,$\times$\,$T_{\rm eff}5$ + 668.
 The numbers of stars in common are 490 and 125;
the correlation coefficients are 0.967 and 0.983, respectively.}}
\end{figure}


\begin{figure}[!t]
\vbox{
\centerline{\psfig{figure=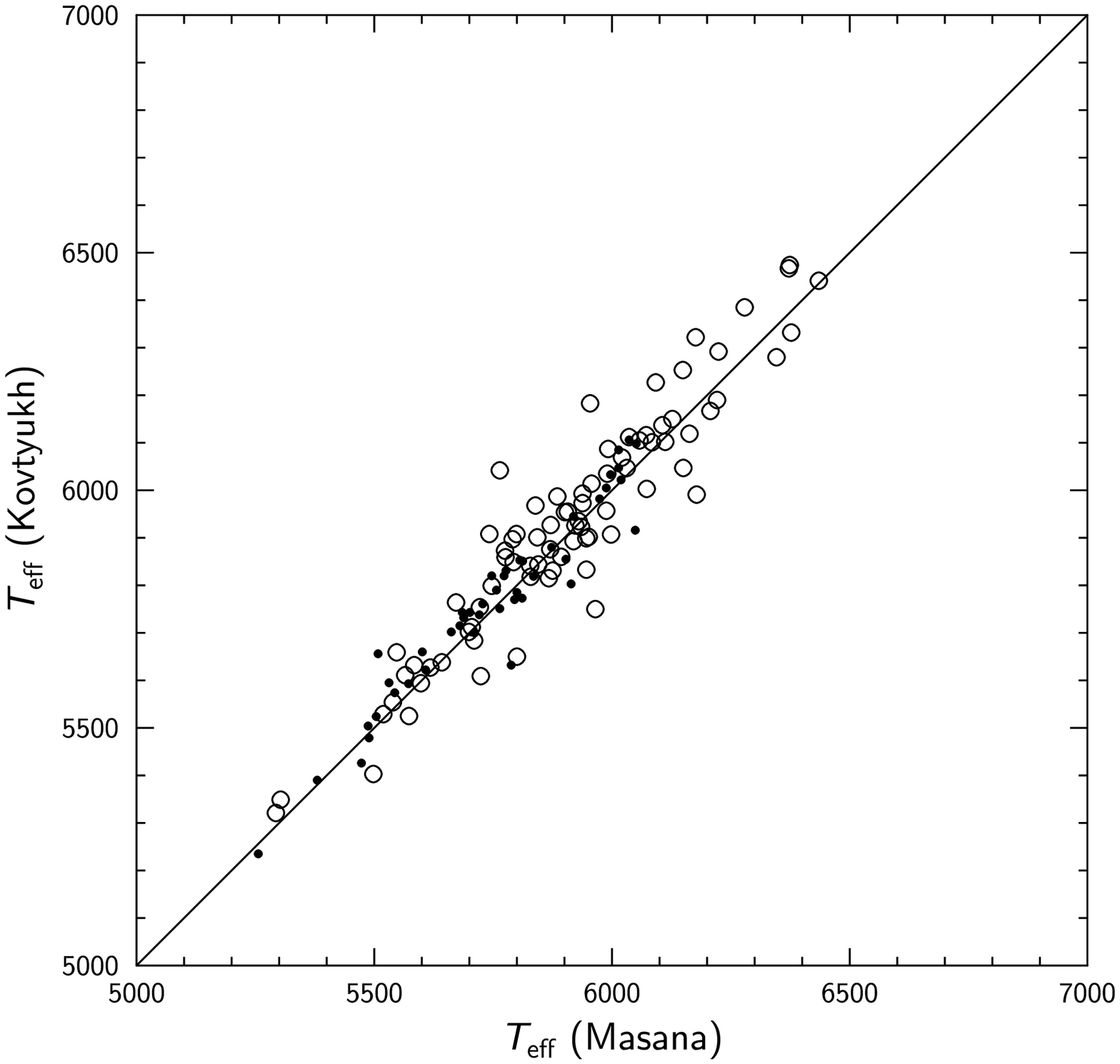,width=80truemm,angle=0,clip=}}
\captionb{2}{Comparison of the effective temperatures from Catalog 1
(Masana et al. 2006) and Catalog 4 (Kovtyukh 2011).  Only the
homogeneous data from Catalog 1 with $\sigma T_{\rm
eff}$\,=\,40--60~K,
$T_{\rm eff}$\,=\,5200--6700 K and [Fe/H]\,$>$\,--1.1 are plotted.
Dots designate the Catalog 4 stars with ${\sigma T_{\rm eff}}<5$ K
(sample standard deviation $s$\,=\,53.9, $N$\,=\,45) and open circles
designate the Catalog 4 stars with ${\sigma T_{\rm eff}}\ge 5$ K
($s$\,=\,82.4, $N$\,=\,81).  } }
\end{figure}

\begin{figure}[!t]
\vbox{
\centerline{\psfig{figure=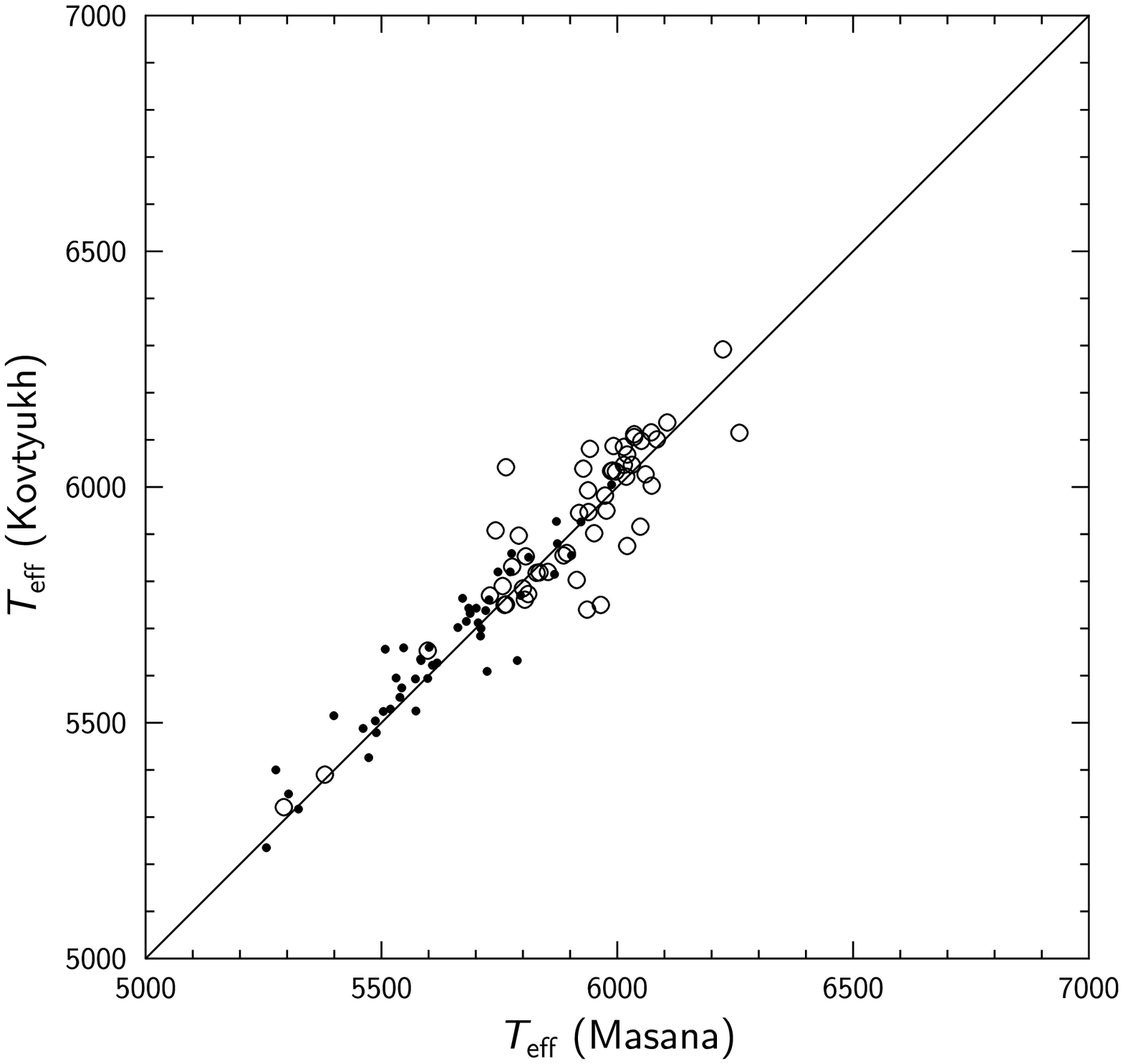,width=80truemm,angle=0,clip=}}
\captionb{3}{The same as in Figure 2 but for other subsamples.  Only the
homogeneous data from Catalog 4 with ${\sigma T_{\rm eff}}$\,=\,2--7 K,
$T_{\rm eff}$\,=\,5200--6700 K and [Fe/H]\,$>$\,--1.1 are plotted.
Dots designate the Catalog 1 stars with ${\sigma T_{\rm eff}}< 50$ K
($s$\,=\,56.0, $N$\,=\,58) and open circles designate the Catalog 1
stars with ${\sigma T_{\rm eff}}\ge 50$ K ($s$\,=\,84.3, $N$\,=\,52).}}
\end{figure}


\begin{figure}[!t]
\vbox{
\centerline{\psfig{figure=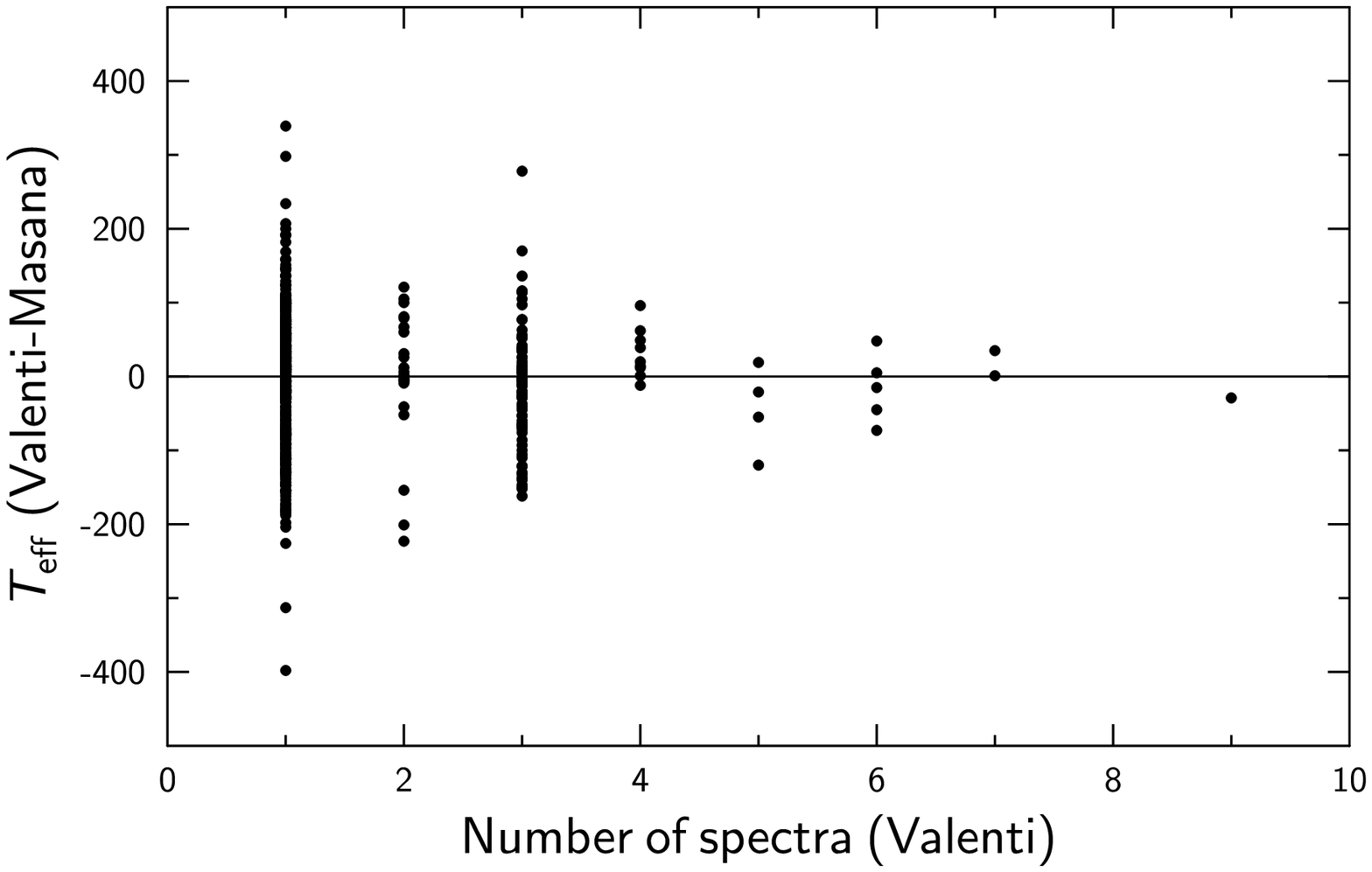,width=80truemm,angle=0,clip=}}
\captionb{4}{Differences of the $T_{\rm eff}$ values between Catalog 2
(Valenti \& Fisher 2005) and Catalog 1 (Masana et al. 2006) versus the
number of the used spectra per star in Catalog 2. The
data of Catalog 1 are limited by the following conditions: $T_{\rm
eff}$\,=\,5200--6700 K and [Fe/H]\,$>$\,--1.1.}}
\end{figure}

In some considered catalogs (1, 4, 5), individual published errors of
$T_{\rm eff}$ values are presented for every star.  We must be cautious
using these errors because some of them may not be reliable, or they are
only internal errors.  These published errors, however, may help us
confine our analysis to more homogeneous subsamples.  The published
errors ${\sigma T_{\rm eff}}$ were plotted versus $T_{\rm eff}$ values
in Fig.~2 of Paper I. The data are rather scattered and the errors, in
general, increase with the temperature.  Hoping to deal with homogeneous
data, in Paper I we selected the most populated data in some narrow
ranges of ${\sigma T_{\rm eff}}$ as subsamples.  Here a
question may arise whether the use of a subsample with smaller published
${\sigma T_{\rm eff}}$ really leads to a smaller scatter in the data
intercomparisons.  To check this, we present two comparisons below.  In
Figure 2, two subsamples from Catalog 4 (one with the published ${\sigma
T_{\rm eff}}<5$ K and another with the published ${\sigma T_{\rm
eff}}\ge 5$ K) are compared with a homogeneous subsample of Catalog 1
explained in the figure.  In Figure 3, two subsamples from Catalog 1
(one with the published ${\sigma T_{\rm eff}}<50$ K and another with the
published ${\sigma T_{\rm eff}}\ge 50$ K) are compared with a
homogeneous subsample of Catalog 4 explained in the figure.  In both
figures, the data are more scattered when the larger published ${\sigma
T_{\rm eff}}$ values are involved.  Therefore, the use of subsamples
with the selected ranges of the published ${\sigma T_{\rm eff}}$
values really
helps us in dealing with more homogeneous data in the comparison.

Catalog 2 (Valenti \& Fisher 2005) contains the $T_{\rm eff}$ values
based on spectral observations (712 stars with only one spectrum and 328
stars with two or more spectra per star), the published error is
 44 K for the single-spectrum stars.  Thus, Catalog 2 should be not
homogeneous with respect to the precision.  To verify, in
Figure 4 we present the differences of the $T_{\rm eff}$ values between
Catalog 1 and Catalog 2 versus the number of used spectra in Catalog 2.
The scatter decreases with the increase of the number of used spectra,
as expected.  Therefore, it would be reasonable to divide Catalog 2 into
subsamples according to the number of the used spectra because we wish
to deal with homogeneous subsamples.

\sectionb{5}{ANALYSIS OF THE SELECTED CATALOGS}

Judging from the results of Paper 1 and according to our discussion in
the previous Section, we have defined the following homogeneous
subsamples of the catalogs with $T_{\rm eff}$\,=\,5200--6700 K for
further analysis:

1. Subsample 1. From Catalog 1 (Masana et al. 2006) we selected the
most populated and homogeneous subsample with ${\sigma
T_{\rm eff}}$\,=\,40--60 K and [Fe/H]\,$>$\,--1.1.

2. Subsamples 2A and 2B.  From Catalog 2 (Valenti \& Fisher 2005) we
 selected two subsamples:  2A -- with two or more spectra per star
and 2B -- with only one spectrum per star.  Only the stars with
[Fe/H]\,$>$\,--1.1 are taken.

3. Subsample 3. From Catalog 3 (Edvardsson et al. 1993) a subsample with
[Fe/H]\,$>$\,--1.1 is  taken.

4. Subsamples 4A and 4B.  From Catalog 4 (Kovtyukh 2011) we selected two
subsamples:  4A -- only with the published errors ${\sigma T_{\rm
eff}}$\,=\,2--7 K and the other (4B) -- with the errors 8--12 K. All
stars in these subsamples are of normal metallicity (with
[Fe/H]\,$>$\,--0.5 where the used calibrations are valid).

5. Subsample 5. From Catalog 5 ( Ramirez \& Melendez 2005) we selected a
subsample with ${\sigma T_{\rm eff}}$\,=\,60--80 K. Since this catalog
does not contain metallicities, [Fe/H] values for this subsample are
taken from other catalogs.  Subsample 5 is not fully independent of
Subsample 1 (some data from 2MASS photometry may be partially used in
both subsamples).

We have analyzed each of these subsamples to determine the errors of
$T_{\rm eff}$ values through data intercomparisons as described in
Section 2. The results are presented in Table 2, where the used
combinations of subsamples are given in the first column.


\begin{table*}
\parbox{120mm}{{\smallbf\ \  Table 2.}\small\baselineskip=9pt\
External ${\sigma T_{\rm eff}}$ for the
selected seven subsamples from five catalogs.}
\scriptsize\tabcolsep=3.5pt
\begin{center}
\begin{tabular}{llllllll}
\tablerule
Combination \hfill & Subs.\,1 \hfill & Subs.\,2A  \hfill & Subs.\,2B
\hfill & Subs.\,3 \hfill & Subs.\,4A \hfill & Subs.\,4B \hfill &Subs.\,5
\\
\tablerule
1-2A-3-4A-5 & 44 & 51 & ~- & 43 & 34 & ~- & 70 \\
1-2A-3-4A  & 49 & 48 & ~- & 42 & 31 & ~- & ~- \\
1-2A-3-4A*  &50$\pm 8$&47$\pm 10$&~-&40$\pm 9$&35$\pm 9$ & ~- & ~- \\
4A-5,2A-5,3-5 & ~- & ~- & ~- & ~- & ~- & ~- & 72$\pm 5$ \\
1-2B,2B-4A  & ~- & ~- & 64$\pm 10$ & ~- & ~- & ~- & ~- \\
1-4B        & ~- & ~- & ~- & ~- & ~- & 63 & ~- \\
\tablerule
Final ${\sigma T_{\rm eff}}$&49&48&64&42&31&63&72\\
\tablerule
Published ${\sigma T_{\rm eff}}$&53$\pm 5$&~-&44&25&4.7$\pm 1$&8.9$\pm
1$&70$\pm 4$\\
\tablerule
\noalign{\vskip2mm}
\multispan6*
{The same data as above but with the use of the technique of Paper I.}\\
\end{tabular}
\end{center}
\end{table*}

Lines 1 and 2 of Table 2 give the errors determined by use of the
combinations of five and four independent subsamples (1-2A-3-4A-5 and
1-2A-3-4A), respectively, by solving linear equations by the least
squares.  For comparison, the line 3 of Table 2 gives the results for
the same subsamples as line 2 but applying a somewhat different
technique (used in Paper I).  We see that the results in lines 2 and 3
are very similar, and in the following analysis we decided to use only
the improved least square technique.  As we suspect that the subsamples
1 and 5 are not completely independent (see above), these two subsamples
have not been used simultaneously in Table 2, except of line 1.

For the remaining subsamples (2B, 4B and 5) we have determined the
errors by processing subsamples in pairs as explained in Section 2
(where one subsample in every pair has already an error estimate from
the second line of Table 2).  The subsamples are sufficiently populated
(minimum there are 32 stars in common in every pair).  For two
subsamples (2B and 5) there are more than one determination for which
the errors of the weighted mean were calculated and presented in Table 2
(with their standard deviations).  The next-to-last line of Table 2
contains our final ${\sigma T_{\rm eff}}$ taken from lines 2 and 4--6.
The last line contains the average published ${\sigma T_{\rm eff}}$ for
each subsample, except of the heterogeneous subsample 2A having various
numbers of the spectra used.

As we can see, our final ${\sigma T_{\rm eff}}$ for the subsamples 1 and
5 do not significantly differ from the corresponding mean published
values given in the last line of Table 2. We may presume that all
sources of errors were properly taken into account calculating the
published errors of the corresponding catalogs.  However, this is not
the case for the subsamples 2A, 2B, 3, 4A and 4B, where our final
${\sigma T_{\rm eff}}$ are significantly larger than the corresponding
published values.  We may suppose that not all sources of errors have
been taken into account in these catalogs.  In the case of subsample 3,
the authors (Edvardsson et al. 1993) have noted indeed that the
published error (25 K) is only due to the errors in the measured
$b$--$y$ values.


\begin{table}
\scriptsize\tabcolsep=3.5pt
\parbox{120mm}{{\smallbf\ \  Table 3.}\small\baselineskip=9pt\
A sample from our mean compiled catalog of the
temperatures.  The involved catalogs are:  Cat.\,1 -- Masana et al.
(2006); Cat.\,2 -- Valenti \& Fisher (2005); Cat.\,3 -- Edvardsson et
al. (1993); Cat.\,4 -- Kovtyukh (2011); Cat.\,5 -- Ramirez \& Melendez
(2005). The numbers of the catalogs used ($n$) and the weighted means of
temperatures with their errors are given in the last columns. The full
catalog will be available from CDS.}
\vspace{-2mm}
\begin{center}
\begin{tabular}{ccccccccccccc}
\tablerule
Number  & Cat.\,1 &${\sigma}$ &Cat.\,2 &${\sigma}$ &Cat.\,3
&${\sigma}$ &Cat.\,4 &${\sigma}$ &Cat.\,5 &${\sigma}$ & $n$ & Wgh. mean
\\
\tablerule
 HD\,\,014938 &    6084 &   62 &    --&   -- & 6164 &   42 &   -- &   -- &   -- &   -- &    2 & 6138 $\pm 34$\\
 HD\,\,015335 &    5977 &   80 & 5898 &   64 & 5857 &   42 & 5950 &   31 &   -- &   -- &    4 & 5919 $\pm 22$\\
 HD\,\,015632 &    5728 &   43 &   -- &   -- &   -- &   -- & 5761 &   31 &   -- &   -- &    2 & 5749 $\pm 25$\\
 HD\,\,016275 &    5834 &   46 & 5848 &   64 &   -- &   -- &   -- &   -- &   -- &   -- &    2 & 5838 $\pm 37$\\
 HD\,\,016397 &    5767 &   48 & 5799 &   64 &   -- &   -- &   -- &   -- &   -- &   -- &    2 & 5778 $\pm 38$\\
 HD\,\,016417 &    5880 &   52 & 5827 &   48 &   -- &   -- &   -- &   -- &   -- &   -- &    2 & 5851 $\pm 35$\\
 HD\,\,016673 &    6224 &   60 &   -- &   -- & 6287 &   42 & 6292 &   31 &   -- &   -- &    3 & 6280 $\pm 23$\\
 HD\,\,017037 &    6189 &   53 & 6203 &   64 &   -- &   -- &   -- &   -- &   -- &   -- &    2 & 6194 $\pm 40$\\
 HD\,\,023596 &      -- &   -- & 5911 &   64 &   -- &   -- & 5931 &   63 & 6055 &   73 &    3 & 5957 $\pm 38$\\
 HD\,\,028005 &      -- &   -- & 5829 &   64 &   -- &   -- & 5977 &   31 &   -- &   -- &    2 & 5948 $\pm 27$\\
 HD\,\,030562 &      -- &   -- & 5943 &   64 & 5886 &   42 & 5836 &   31 & 5877 &   69 &    4 & 5866 $\pm 22$\\
 HD\,\,030649 &      -- &   -- & 5789 &   64 & 5736 &   42 &   -- &   -- &   -- &   -- &    2 & 5751 $\pm 35$\\
 HD\,\,032963 &      -- &   -- & 5775 &   64 &   -- &   -- & 5741 &   31 &   -- &   -- &    2 & 5747 $\pm 27$\\
 HD\,\,006434 &      -- &   -- &   -- &   -- & 5813 &   42 &   -- &   -- & 5842 &   70 &    2 & 5820 $\pm 36$\\
 HD\,\,010307 &      -- &   -- &   -- &   -- & 5898 &   42 & 5891 &   31 & 5965 &   73 &    3 & 5900 $\pm 23$\\
 HD\,\,014214 &      -- &   -- &   -- &   -- & 6045 &   42 & 6035 &   31 & 5977 &   74 &    3 & 6032 $\pm 23$\\
 HD\,\,038393 &      -- &   -- &   -- &   -- & 6398 &   42 & 6388 &   31 & 6327 &   73 &    3 & 6384 $\pm 23$\\
 HD\,\,041330 &      -- &   -- &   -- &   -- & 5917 &   42 & 5933 &   31 &   -- &   -- &    2 & 5927 $\pm 24$\\
 HD\,\,043318 &      -- &   -- &   -- &   -- & 6347 &   42 & 6340 &   31 &   -- &   -- &    2 & 6342 $\pm 24$\\
\tablerule
\end{tabular}\end{center}
\vspace{-4mm}
\end{table}

\sectionb{6}{THE MEAN COMPILED CATALOG}

We have produced a mean compiled homogenized catalog of $T_{\rm eff}$
values for stars from the five selected catalogs described above,
where we use our estimates of ${\sigma T_{\rm eff}}$ for defining
weights in averaging the data, as explained in Section
2. The compilation of the mean catalog involves the following steps:

1. The data are used only if they are available at a minimum in two
catalogs.

2. We consider only the stars with [Fe/H]\,$>$\,--1.1 in Catalogs 1,
2, 3 and 5, with [Fe/H]\,$>$\,--0.5 in Catalog 4, and having $T_{\rm
eff}$ = 5200--6700 K in all catalogs.

3. The $T_{\rm eff}$ values are averaged with the weights based on our
final ${\sigma T_{\rm eff}}$ results given in Table 2, but with some
modifications.  In the case of Catalogs 1 and 5, our final ${\sigma
T_{\rm eff}}$ values for the respective subsamples are similar to the
mean published ${\sigma T_{\rm eff}}$ for these subsamples (see the
previous section).  It may be reasonable to suppose that such similarity
exists for any other subsample of these catalogs.  Thus we decided to
use the published individual ${\sigma T_{\rm eff}}$ values for defining
the weights for
all stars of Catalogs 1 and 5, instead of our final ${\sigma T_{\rm
eff}}$ values.

The mean weighted $T_{\rm eff}$ values and their errors for the stars in
the compiled catalog were calculated using Eqs.  (1)--(3), the total
number of stars included is 800.


\begin{figure}[!t]
\vbox{
\centerline{\psfig{figure=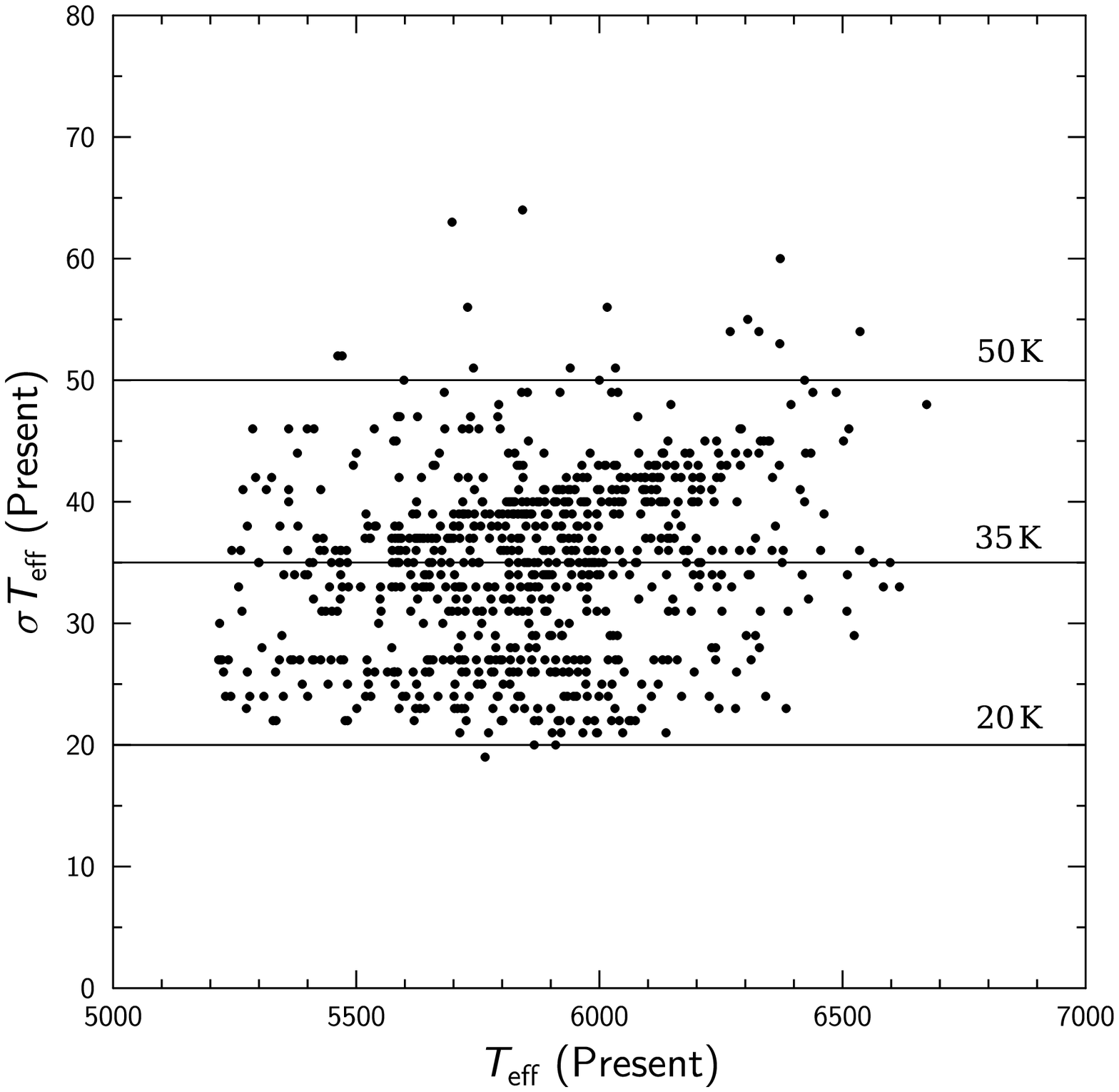,width=80truemm,angle=0,clip=}}
\captionb{5}{The ${\sigma T_{\rm eff}}$ values
plotted versus $T_{\rm eff}$, the data are from the present catalog.
The homogeneous
subsamples A and B with the errors of $T_{\rm eff}$ within the
indicated intervals were selected for further analysis. Subsample A:
20\,$\leq \sigma T_{\rm eff}$\,$<$\,35 K; mean = 28\,$\pm$\,4~K; $N$ =
355. Subsample B:
35\,$\leq \sigma T_{\rm eff}$\,$<$\,50 K; mean = 40\,$\pm$\,4~K; $N$ =
429.}}
\end{figure}

As an illustration, a sample from the mean temperature catalog is
presented in Table 3.
In Figure 5, the $T_{\rm eff}$ values of the present catalog are
compared with their errors.  Two homogeneous subsamples A and B were
selected with ${\sigma T_{\rm eff}}$ between the horizontal lines; the
mean values, the standard deviations and the numbers of the subsample
stars are given in the figure.  These subsamples will be used for
analysis of other catalogs in the next section.  In Figure 5 a general
increase of the errors with $T_{\rm eff}$ may be noted, it is a typical
feature of all considered catalogs.  The mean errors of $T_{\rm eff}$
values in our mean catalog are significantly smaller than the
corresponding individual errors given in Table 3.

\begin{figure}[!t]
\vbox{
\centerline{\psfig{figure=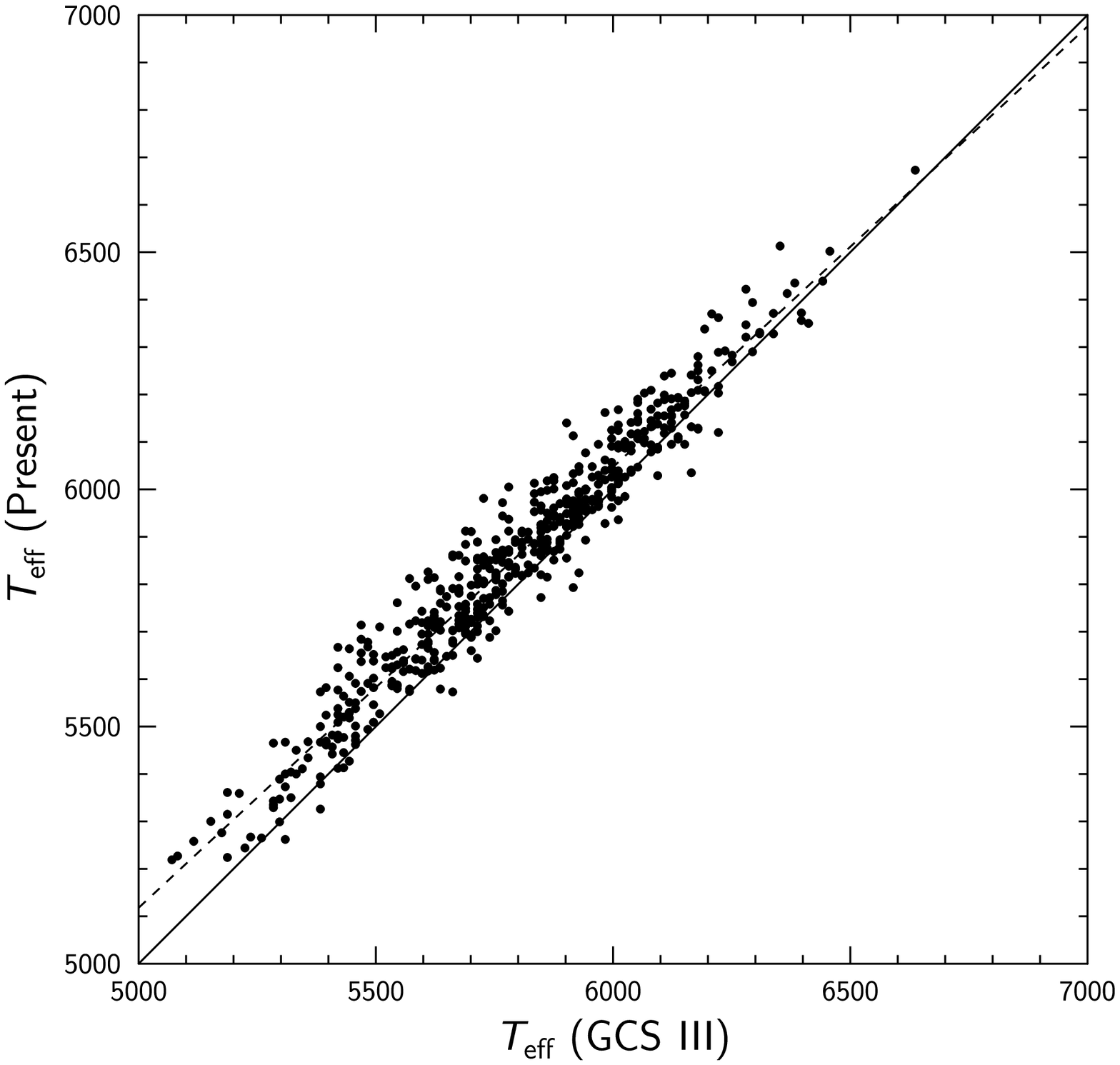,width=80truemm,angle=0,clip=}}
\captionb{6}{ Comparison of the effective temperatures between the GCS
III catalog (Holmberg et al. 2009) and the present catalog.
The transformation equation  (broken line) is
$T_{\rm eff}$ = 0.9293\,$\times$\,$T_{\rm eff}$\,(GCS\,III) + 471.
 The number of stars in common is 532;
the correlation coefficient is 0.971.
}}
\end{figure}

\begin{figure}[!t]
\vbox{
\centerline{\psfig{figure=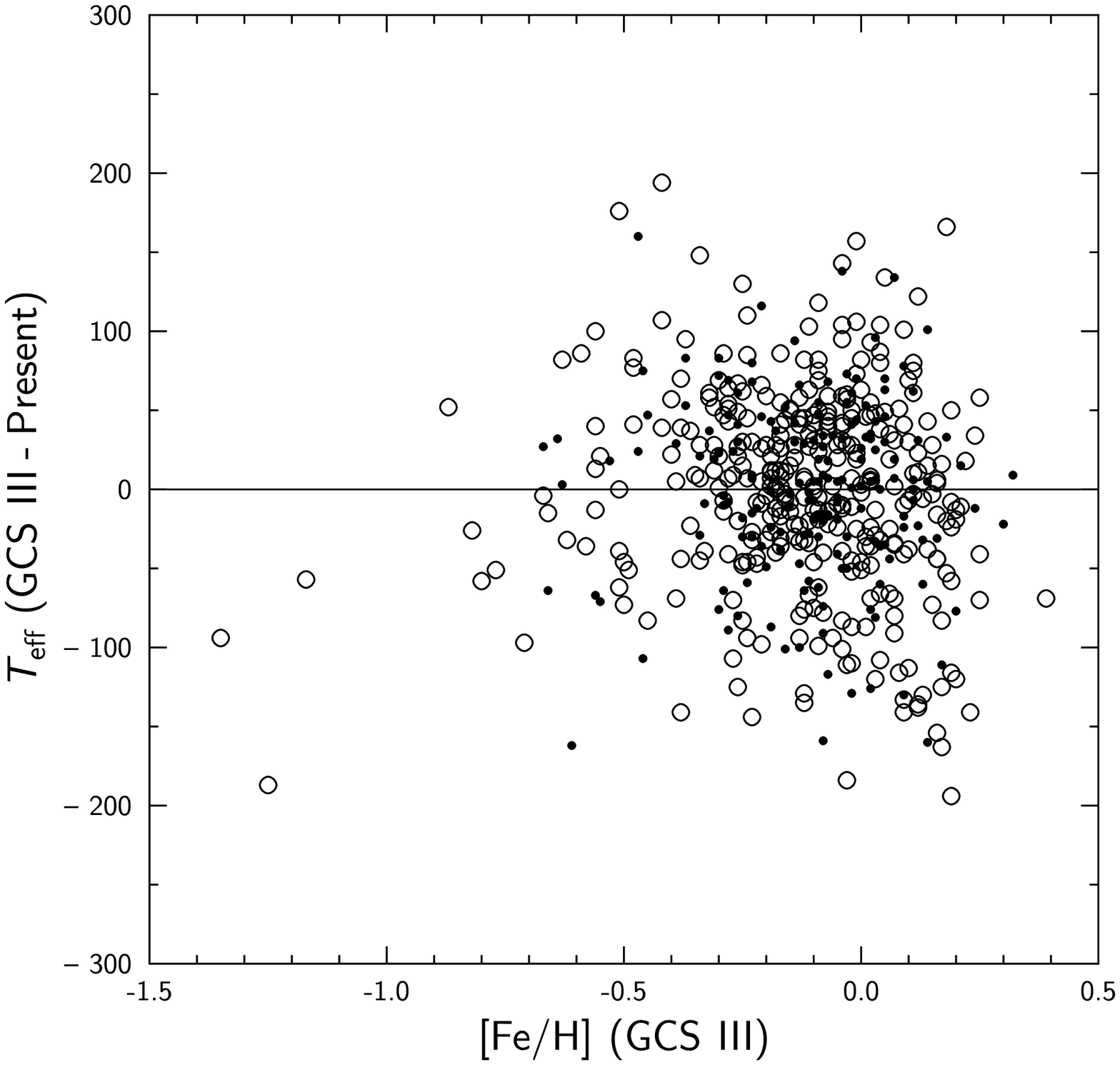,width=80truemm,angle=0,clip=}}
\captionb{7}{Differences of  $T_{\rm eff}$  between the  GCS III
catalog and the subsamples A (dots) and B (open circles) of the present
catalog versus [Fe/H] from GCS III.}
}
\vspace{8mm}
\end{figure}

\sectionb{7}{ERROR ESTIMATES OF THE EFFECTIVE TEMPERATURES IN\\
INDEPENDENT CATALOGS WITH THE USE OF THE PRESENT DATA}
\vskip2mm

Two extensive homogeneous subsamples A and B defined in Figure 5, having
the reliable mean external errors of $T_{\rm eff}$ (28 K and 40 K,
respectively) may be used to estimate the external errors of the
effective temperatures in some other independent catalogs from the
variances of the data differences (see Eqs.  (8) and (9)).  Some
examples of such approach are given below.

\subsectionb{7.1}{The catalog GCS III}

A magnitude complete sample of 16\,682 F--G dwarfs in the solar vicinity
was presented and analyzed in the Geneva-Copenhagen survey (GCS III
catalog, Holmberg et al. 2009), where the $T_{\rm eff}$ (as well
as some other parameters) values were derived from photometry in the
Str\"omgren system with the use of recent calibrations.  The GCS III
catalog does not contain errors of the given temperatures, although
these errors may be important in many applications, especially for using
them as weights in compilation of new catalogs.

Since GCS III and the present catalog contain 130 stars in common, the
same photometric data could be used in both catalogs.  To provide the
use of only independent measurements, we modified GCS III catalog by
excluding the mentioned 130 stars from consideration.  One more
modification of the GCS III catalog was the reduction of its
temperatures to the system of the present catalog by a linear equation
(Figure 6).  Below we will consider only this modified GCS III catalog.

First, we analyzed possible systematic trends of $T_{\rm eff}$
differences between the GCS III Catalog and the subsamples A and B of
the present catalog with some physical parameters.  Figure 7 shows that
there is no any trend of these differences with [Fe/H], at least for
[Fe/H]\,$>$\,--1.1.

\begin{figure}[!t]
\vbox{
\centerline{\psfig{figure=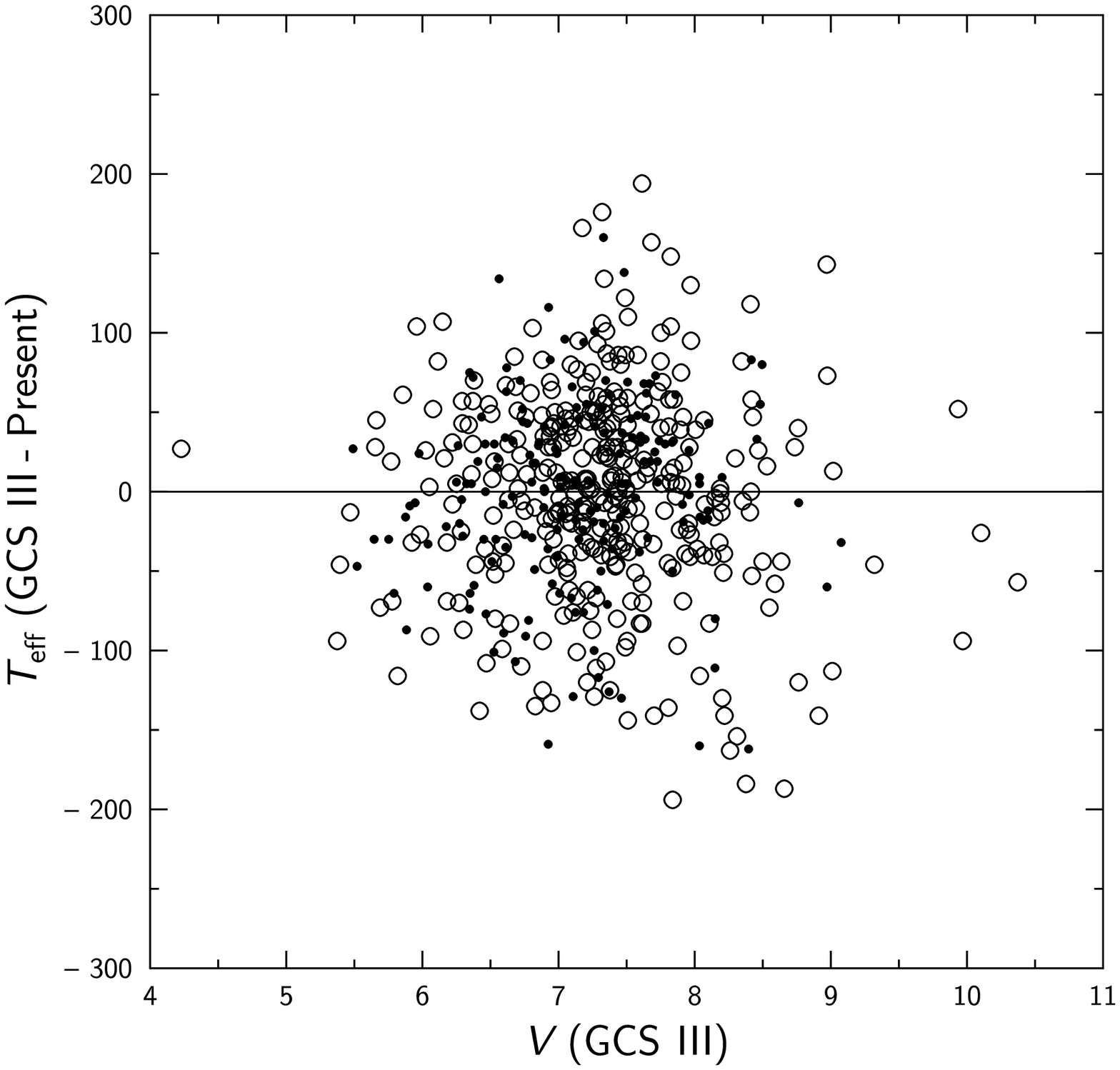,width=80truemm,angle=0,clip=}}
\captionb{8}{Differences of $T_{\rm eff}$ between the GCS III catalog
and our subsamples A and B versus $V$ magnitude from GCS III.
Designations are the same as in Figure 7.}
}
\end{figure}

Because a magnitude-complete sample of stars in the GCS III catalog is
used, we may suspect that the $T_{\rm eff}$ differences depend on
magnitude.  In Figure 8 the $T_{\rm eff}$ differences (GCS III minus the
present catalog) are plotted against $V$ magnitudes taken from GCS III
(reddening effects on $V$ are small and can be neglected).  We see that
the less reliable data for the stars of our subsample B are more
scattered, as expected.  We see also that the scatter for both
subsamples increases with $V$ due to lower accuracy of the GCS III data
for fainter stars.

To deal with more homogeneous data, we divided the GCS III catalog into
two groups:  Group 1 with $V<7.0$ and Group 2 with $V\ge 7.0$.  Then we
calculated the variances of the $T_{\rm eff}$ differences (GCS III minus
the present catalog) for the stars in these groups (as well as for data
with all $V$) for the subsamples A and B. From these variances, with the
use of the external errors of $T_{\rm eff}$ for the subsamples A and B,
we calculated the external errors of $T_{\rm eff}$ for every GCS III
group.  The results are presented in Table 4. We see that the derived
errors of $T_{\rm eff}$ are very consistent for every GCS III group, the
errors for fainter stars being larger, as expected.  The ${\sigma T_{\rm
eff}}$ for the brightest stars (Group 1) from GCS III (44 K for the mean
$V$\,=\,6.49) is very similar to our $T_{\rm eff}$ error estimate (42 K)
obtained for the catalog of Edvardsson et al.  (1993) in Table 2, where
independent $b$--$y$ photometry has been used with the mean
$V$\,=\,5.79.  We consider that the GCS III catalog may serve as one
more data source in producing homogenized compiled catalogs with the use
of error estimates from Table 4.


\vspace{3mm}
\begin{table*}
\parbox{120mm}{{\smallbf\ \  Table 4.}\small\baselineskip=9pt\
Determination of the external ${\sigma T_{\rm eff}}$ for
some groups of the  GCS III catalog
through their intercomparisons with the subsamples A and B of the
present catalog.}
\begin{center}
\begin{tabular}{llll}
\tablerule
\noalign{Group 1 of GCS III with $V<7.0$ (mean $V$\,=\,6.49)}
\tablerule
& $N$ & St. deviation & ${\sigma T_{\rm eff}}$\\
\tablerule
With subsample A & 76 & 53.0 & 45 K  \\
With subsample B & 97 & 59.7 & 44 K \\
&&&
Weighted mean ${\sigma T_{\rm eff}}$\,=\,44 K\\
\tablerule
\tablerule
\noalign{Group 2 of GCS III with $V\ge 7.0$ (mean $V$\,=\,7.64)}
\tablerule
& $N$ & St. deviation & ${\sigma T_{\rm eff}}$\\
\tablerule
With subsample A & 101 & 59.9 & 53 K  \\
With subsample B & 247 & 67.4 & 54 K \\
&&& Weighted mean ${\sigma T_{\rm eff}}$\,=\,54 K\\
\tablerule
\tablerule
\noalign{The GCS III catalog with all $V$ (mean $V$=7.26)}
\tablerule
& $N$ & St. deviation & ${\sigma T_{\rm eff}}$\\
\tablerule
With subsample A & 177 & 57.0 & 50 K  \\
With subsample B & 344 & 65.2 & 52 K \\
&&& Weighted mean ${\sigma T_{\rm eff}}$\,=\,51 K\\
\tablerule
\end{tabular}\end{center}
\noindent\end{table*}

\subsectionb{7.2}{The Fuhrmann catalog}

This sample (Fuhrmann 1998) contains about 50 nearby F and G stars,
dwarfs and subgiants of the Galactic disk and halo.  Effective
temperatures were determined from fits of the synthetic spectra to
wings of the Balmer lines.  This sample is relatively small but
important for our analysis because the used data are completely
independent from the present catalog, and the analyzed wings of the
Balmer lines are very sensitive to the temperature (Schmidt 1972).


\begin{figure}[!t]
\vbox{
\centerline{\psfig{figure=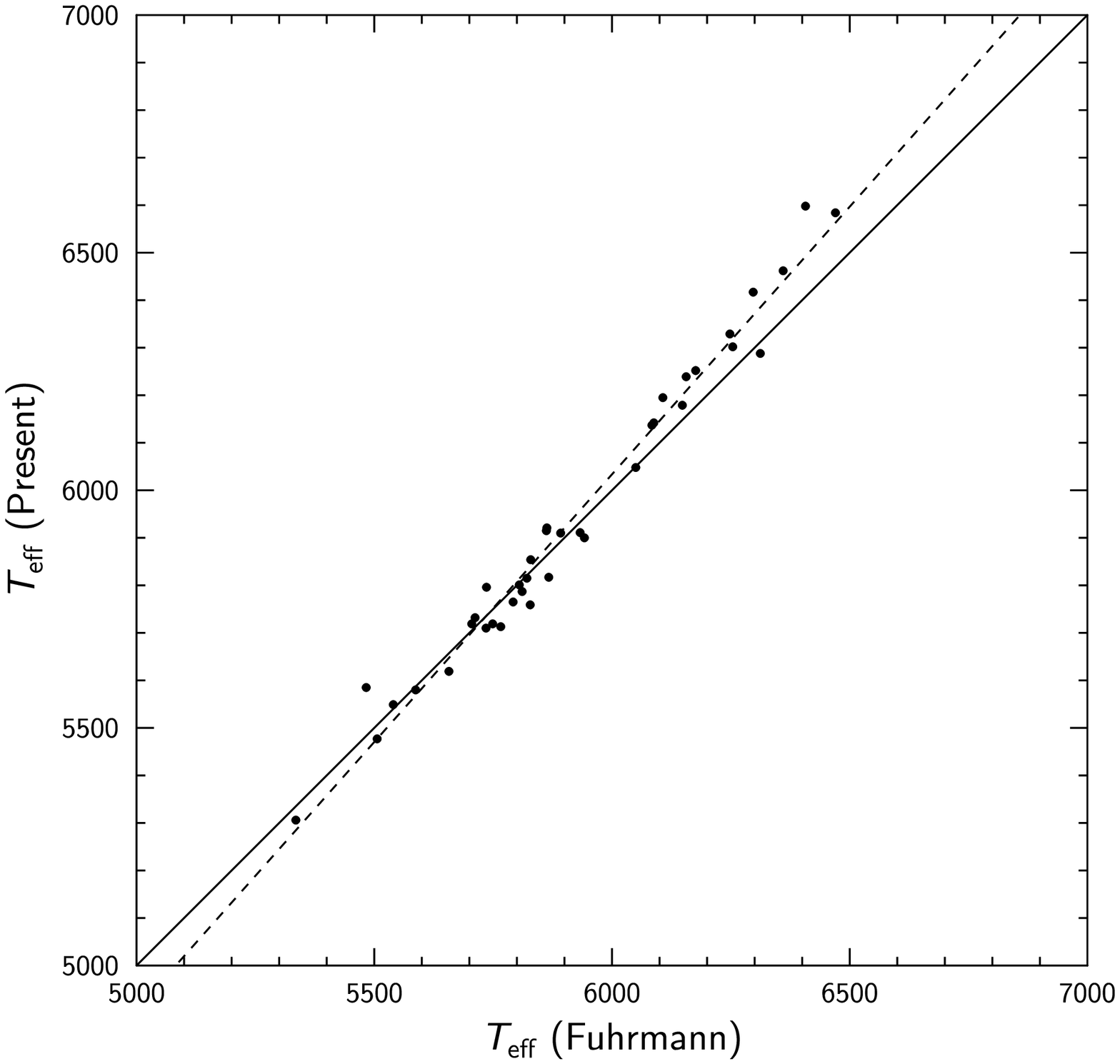,width=80truemm,angle=0,clip=}}
\captionb{9} {The same as in Figure 6 but for the Fuhrmann (1998)
catalog.
The transformation equation  (broken line) is
$T_{\rm eff}$ = 1.1266\,$\times$\,$T_{\rm eff}$\,(Fuhr) -- 726.
 The numbers of stars in common is 38;
the correlation coefficient is 0.989.
}}
\end{figure}

\begin{figure}[!t]
\vbox{
\centerline{\psfig{figure=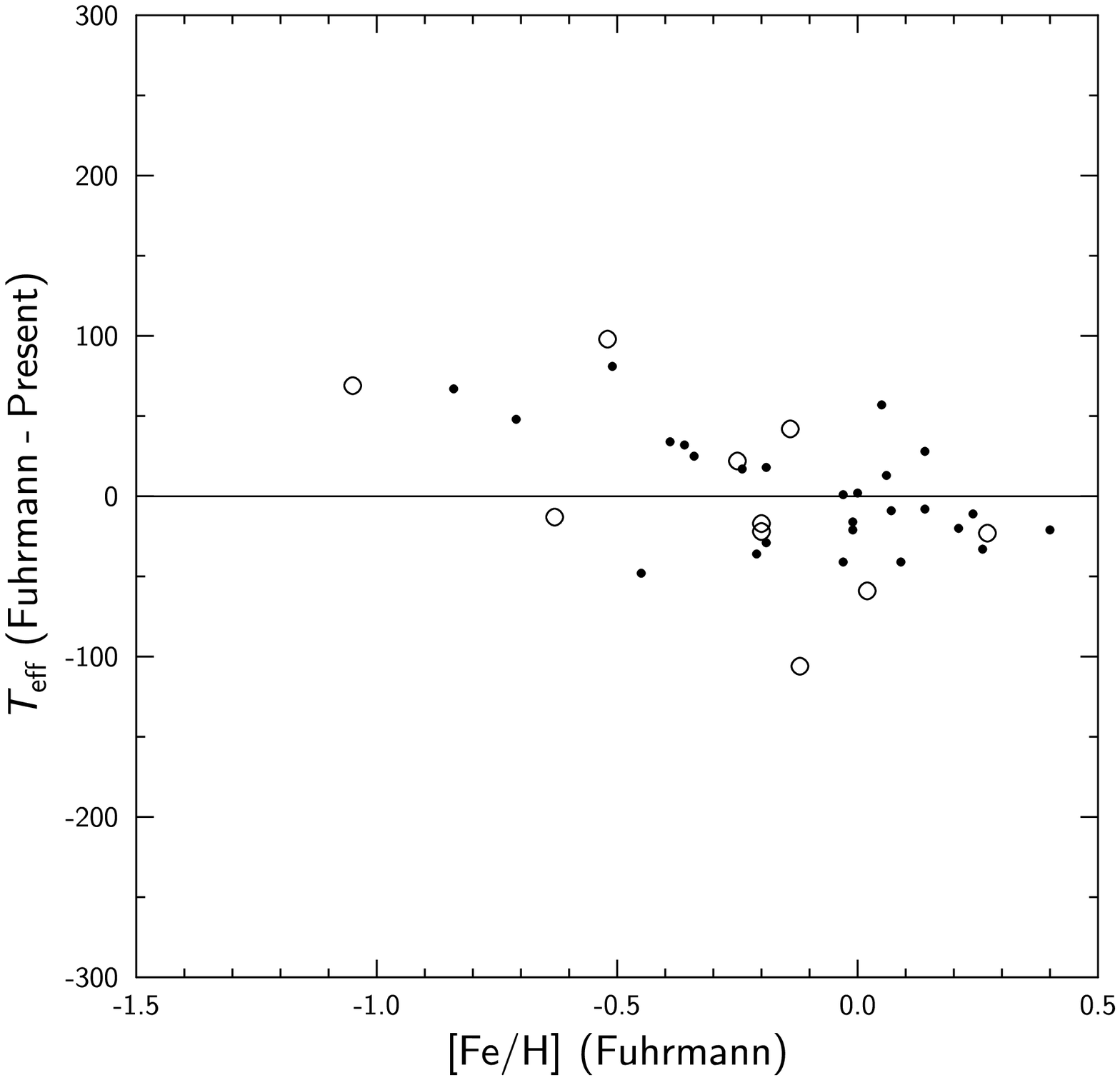,width=80truemm,angle=0,clip=}}
\captionc{10}{The same as in Figure 7 but for the Fuhrmann (1998)
catalog.}  }
\vspace{11mm}
\end{figure}

One necessary modification of the Fuhrmann catalog was the reduction of
its $T_{\rm eff}$ to the system of the present catalog (Figure 9).
Figure 10 shows that the $T_{\rm eff}$ differences (the reduced Fuhrmann
catalog minus the present catalog) exhibit no trend with [Fe/H].  The
data from the subsample B are more scattered, as expected.

Then we calculated the variances of the same data differences for the
subsamples A and B:  the standard deviations are 35.4 ($N$\,=\,26) and
60.4 ($N$\,=\,10), respectively.  From these variances, with the use of
the appropriate external errors of $T_{\rm eff}$ for the subsamples A
and B, we calculated the external errors of $T_{\rm eff}$ for the
Fuhrmann data:  21~K with the use of subsample A and 45~K with the use
of subsample B, the weighted average value being 28 K.

Considering the external error estimates of all the catalogs
investigated in the present paper (Table 2 and Section 7), we conclude
that the lowest errors ${\sigma T_{\rm eff}}$ are for the Fuhrmann
(1998) catalog (28~K) and for subsample 4A of the Kovtyukh (2011)
catalog (31~K).  For the Edvardsson et al.  (1993) and the GCS III
catalogs ($V<7.0$) $\sigma T_{\rm eff}$ are 42 and 44 K, respectively,
for the other catalogs they are $\sim$\,50~K or somewhat larger.

\sectionb{8}{CONCLUSIONS}

We have analyzed some selected catalogs of stellar $T_{\rm eff}$ values,
estimated the errors of these catalogs (and/or of some their subsamples)
from data intercomparisons, and produced the compiled homogenized
catalog based on these data.  The present catalog may be used as a
source of homogeneous $T_{\rm eff}$ values, together with their errors.
The homogeneous subsamples extracted from the present catalog can be
used as the comparison data sources for estimating external errors of
$T_{\rm eff}$ in other catalogs through the data
comparisons.  The same approach may be applied also for treatment
of other data types (gravities, metallicities, magnitudes, color
indices, etc.).  In future, we hope to produce more spacious homogeneous
samples of stars with reliable astrophysical parameters important for
studies of the Galactic structure and evolution, using the observations
in large photometric and spectral surveys.


\thanks{ The authors are grateful to J. Pelt, V. Marsakov and N.
Kharchenko for useful suggestions and fruitful discussions.  We thank an
anonymous referee whose important comments improved the quality of this
paper.  Support from the Estonian Science Foundation (grant No. 7765) is
gratefully acknowledged.  This reasearch has used the SIMBAD database
operated at CDS, Strasbourg, France.}

\References

\refb Borkova~T.~V., Marsakov~V.~A. 2005, Astronomy Reports, 49, 405

\refb Chun~M. S., Freeman~K. C. 1978, AJ, 83, 376

\refb Cenarro~A.  J., Peletier~R.  F., Sanchez-Blazquez~P. et al. 2007,
MNRAS, 374, 664

\refb Ducourant~C., Le Campion~J.  F., Rapaport~M. et al. 2006, A\&A,
448, 1235

\refb Edvardsson~B., Andersen~J., Gustafsson~B. et al. 1993, A\&A, 275,
101

\refb Evans~D.  W., Irwin~M.  J., Helmer~L. 2002, A\&A, 395, 347

\refb Fuhrmann~K. 1998, A\&A, 338, 161

\refb Holmberg~J., Nordstr\"om~B., Andersen~J. 2009, A\&A, 501, 941

\refb Kovtyukh~V.~V. 2011, private communication

\refb Malyuto~V., Shvelidze~T. 2008, Baltic Astronomy, 17, 373 (Paper I)

\refb Malyuto~V. 1994, A\&AS, 108, 441

\refb Masana~E., Jordi~C., Ribas~I. 2006, A\&A, 450, 735

\refb Masana~E., 2008, private communication

\refb Ramirez~I., Melendez~J. 2005, ApJ, 626, 446

\refb Schmidt~E. G. 1972, ApJ, 174, 595

\refb Soubiran~C, Le Campion~J.-F., Cayrel de Strobel~G., Caillo~A.
2010, A\&A, 515, A111

\refb Taylor~B. J. 2005, ApJS, 161, 444

\refb Valenti~J.~A., Fisher~D.~A. 2005, ApJS, 159, 141

\end{document}